\documentclass[conference]{IEEEtran}

\ifCLASSINFOpdf
   \usepackage[pdftex]{graphicx}
\else
   \usepackage[dvips]{graphicx}
\fi
\usepackage[cmex10]{amsmath}
\usepackage{url}

\usepackage[]{fancyhdr} %
\newcommand{\changefont}{\fontsize{9}{9}\selectfont}
\usepackage{xcolor}
\usepackage{hhline}
\usepackage{multirow}
\usepackage{graphicx}
\usepackage{hyperref}
\hypersetup{
	colorlinks=false,
	pdfborder = {0 0 0}
}

\IEEEoverridecommandlockouts
\begin{document}

%
\title{Method to Build Equivalent Models of Microgrids for RMS Dynamic Simulation of Power Systems}



 
 \author{\IEEEauthorblockN{Rodrigo A. Ramos\IEEEauthorrefmark{1},
		Ahda P. Grilo-Pavani\IEEEauthorrefmark{2},
		Artur B. Piardi\IEEEauthorrefmark{1} and 
		Tatiane C. C. Fernandes\IEEEauthorrefmark{3}}
	\IEEEauthorblockA{\IEEEauthorrefmark{1} 
		University of Sao Paulo,
		Sao Carlos, Brazil\\ 
		Emails: \href{mailto:rodrigo.ramos@ieee.org}{rodrigo.ramos@ieee.org} and \href{mailto:artur.b.piardi@gmail.com}{artur.b.piardi@gmail.com}
	}
	\IEEEauthorblockA{\IEEEauthorrefmark{2} 
		Federal University of ABC,
		Santo Andre, Brazil\\ 
		Email: \href{mailto:ahda.pavani@ufabc.edu.br}{ahda.pavani@ufabc.edu.br}
	}
	\IEEEauthorblockA{\IEEEauthorrefmark{3}
		Federal University of Sao Carlos,
		Sao Carlos, Brazil\\
		Email: \href{mailto:tatianefernandes@ufscar.br}{tatianefernandes@ufscar.br}
}}





\maketitle
\thispagestyle{fancy}
\pagestyle{fancy}


\begin{abstract}
The high penetration of distributed renewable energy resources in power systems has changed their dynamic behavior, not only at the distribution level but also at the transmission levels. For analyses performed in this new reality of interconnected systems, a suitable equivalent model is required to represent the active dynamics of distribution systems. In this context, this paper proposes the application of a gray-box method to obtain an appropriate equivalent model for active distribution networks. From data measured at the point of common coupling, a trajectory sensitivity analysis is carried out to select the most important parameters of this equivalent model, which are then estimated by an evolutionary algorithm. The results show that the application of the sensitivity analysis can improve the quality of the parameter estimation process (since it focuses only on relevant parameters), enabling an efficient tuning of an equivalent ADN model. 	
\end{abstract}

\begin{IEEEkeywords}
Distributed Energy Resources; Microgrids; Active Distribution Networks; Equivalent Models; Reduced-order Models; Gray-Box Techniques.
\end{IEEEkeywords}


%
\IEEEpeerreviewmaketitle

\section{Introduction}

Microgrids (MGs) are emerging rapidly in medium and low voltage distribution systems, as they allow the efficient integration of distributed energy resources (DER), especially based on renewables, enabling a higher penetration of DERs in distribution systems \cite{Olivares2014}. This large-scale integration of DERs is changing the behavior of the distribution systems from a passive to an active network. As a consequence, the traditional representation of distribution systems by lumped load models in studies of power systems dynamic performance is no longer suitable for several distribution systems. This scenario calls for new modelling strategies to properly identify the dynamic response of Active Distribution Networks (ADNs), since these networks have different characteristics and employ different technologies from the conventional generators connected to bulk power systems \cite{Milanovic2013,Zaker2019,resende}. 

As the ADNs can be composed of several MGs, the first step to model an ADN is modeling a microgrid, and then the same technique can be used to model a whole ADN for dynamic studies. Although using the full-scale model of the grid would be the first choice, it is not an option due to required high computational burden. Besides, there is a high complexity involved in obtaining data from all components present in a microgrid. Hence, equivalent aggregated MG models have been proposed in the literature. 

As discussed in \cite{resende}, conventional dynamic equivalence techniques are not suitable to derive MG equivalents. To overcome this issue, nonlinear system identification theory has been exploited for this purpose. These approaches use measured data to build mathematical models of the equivalent system and do not require a detailed knowledge of the architecture and the composition of the MG. The method can be based on a black-box or a gray-box approach, depending on the level of available knowledge and physical insight about the system. Black-box approaches do not assume any pre-established model and do not depend on information about the system architecture. In this type of approach, techniques such as artificial neural networks (ANN) \cite{erlich2004a,erlich2004b} and Prony-based method \cite{pana2015} have been employed. However, for transmission system operators, the use of black-box approaches is not well accepted since the model is not physically interpretable and difficult to be implemented with the built-in models in the commercial tools available for dynamic simulation. 

When some physical knowledge about the system structure is available, a model structure can be selected to represent the system, i.e., it is possible to adopt a gray-box approach. A gray box model is practical to be integrated in dynamic simulation tools since it uses models of typical components of the system. In the literature, it is possible to find some gray-box models for representing ADN or MGs. In \cite{Milanovic2013}, it is proposed a gray-box model for ADN comprising a model of a  converter-interfaced synchronous generation and a composite load model. 
In the method used, measurements of active and reactive powers, voltage, and frequency are collected at the point of common coupling (PCC). The measurements are used to estimate the parameters of the proposed model.
The model present good results, but as its generator equivalent is a converter-interfaced synchronous generation, it is limited to represent MG, especially the ones with battery energy storage systems. 
In \cite{chasp2018}, the authors propose a gray-box aggregated equivalent model, designed to represent low and medium-voltage with inverter-based generation (IBG) and loads into two lumped equivalent IBGs. In this case, the Differential Evolution algorithm is used to estimate the parameters of the equivalent system. The same equivalent is used in \cite{chasp2020}, but to address the uncertainty of the operation point, they used the average of the randomized responses of Monte-Carlo simulations to identify the parameters of the equivalent. 

An equivalent model for grid-connected microgrids including a synchronous generator, a voltage source converter (VSC), and a static load model is proposed in \cite{Zaker2019}. Although the model of generation is complete, the load model does not allow representing dynamic loads, which in MGs may present a high impact on the system response. An adaptive structure is proposed in \cite{conte2019}, using asynchronous machine, a ZIP load and a synchronous generator (SG), allowing that the machines can be connected to the grid by a converter. A gray-box model composed by a power converter, a synchronous generation unit; and a composite load model is proposed in \cite{nuno2020}. The model parameters are estimated by an evolutionary particle swarm optimization algorithm. 

An analyses of the methods available in the literature allows verifying that an equivalent model composed by a VSC, a synchronous generator, and a composite load can suitably represent the equivalent of an ADN, as well as a grid-connected MG. However, considering references that employ this type of equivalent (e.g. \cite{Zaker2019, nuno2020}), it is possible to note that there is a large number of parameters to be identified and there is no procedure to limit the range of the search in the parameter space. When combined, these two features can result in numerical problems, slowing down the search or even making it computationally infeasible. 

In this context, the main contribution of this paper is the proposal of a gray-box equivalent model (employing composed a VSC, a synchronous generator, and a composite load) and the use of a Trajectory Sensitivity Analysis (TSA) to select the parameters of it that must be estimated. The use of TSA allows selecting the parameters that have influence on the dynamic response required for stability studies. 
Then, a Differential Evolution (DE) algorithm is applied to identify the set of parameters selected by the TSA (a smaller subset of the set of all equivalent model parameters, which reduces the numerical issues mentioned in the previous paragraph). 

This paper is organized as follows: Section \ref{sec:equivalent_model}  presents the proposed equivalent model structure and its components. The full method proposed to identify the parameters of the equivalent model is explained in Section \ref{sec:method}, and a case study is shown in Section \ref{sec:results}. Finaly, Section \ref{sec:conclusions} concludes the paper.


\section{Proposed Equivalent Model}
\label{sec:equivalent_model}

\begin{figure}[ht]
	\centering
	\includegraphics[width=0.25\textwidth]{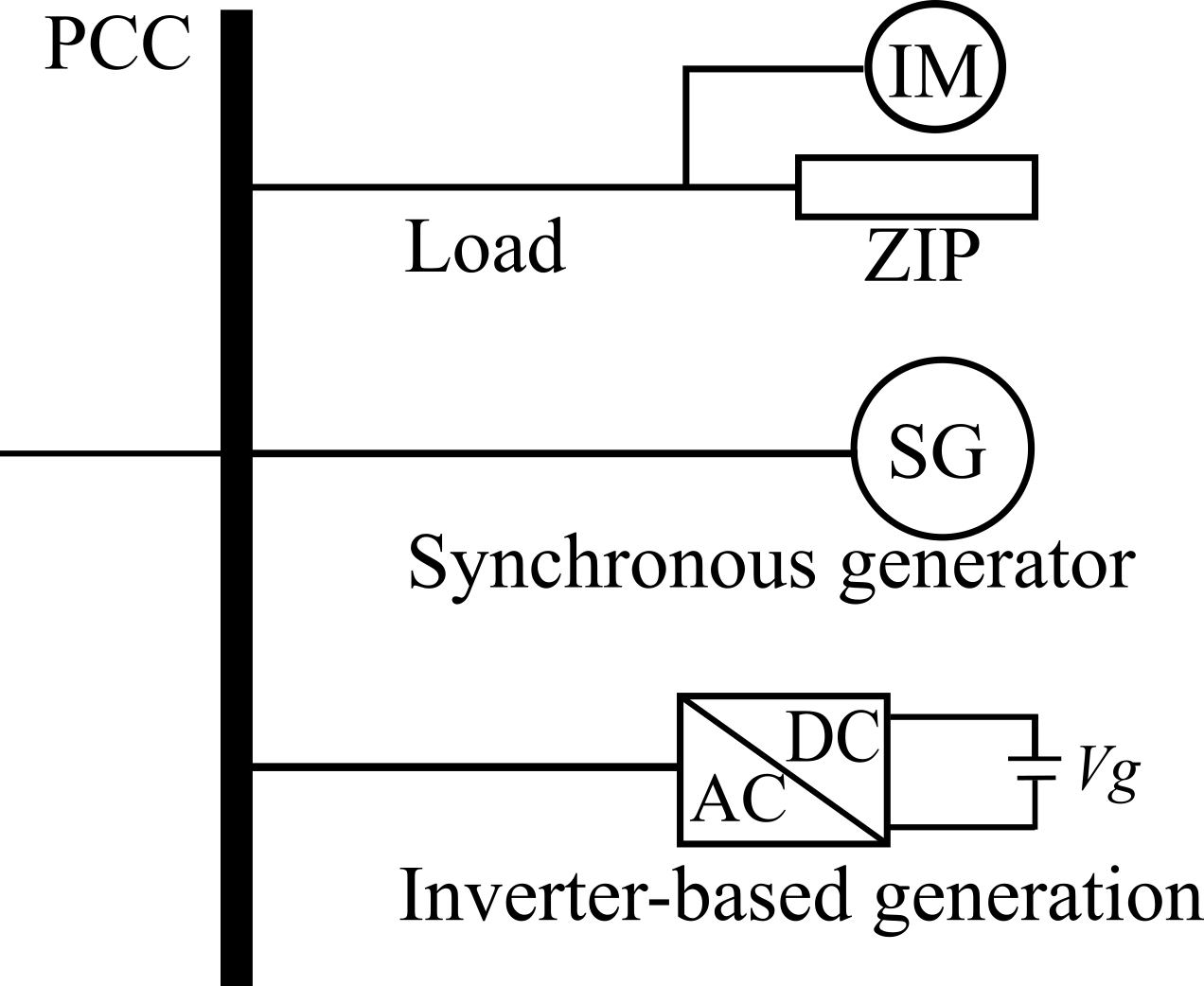}
	\caption{Equivalent microgrid model.}
	\label{fig:equiv_model}
\end{figure}

In this work, the focus is the medium-voltage MGs, which are usually part of a distribution system. Typically, the generation connected to the distribution system comprises synchronous generation units and converter-interfaced generation units, used for wind and PV generation, as well as energy storage systems (ESS). Thus, it is proposed the use of the equivalent model structure presented in Fig. \ref{fig:equiv_model}. The proposed equivalent model is comprised of a synchronous generator unit, a composite load, and a voltage source converter. These components are connected in parallel to the point of common coupling (PCC) and may represent a medium-voltage MG, as well as an ADN since it presents the main elements that dynamically contribute to the behaviour of the overvall MG or ADN. 

The main idea of building an equivalent is to use it for representing the MG during a disturbance, that is, when connected to the main grid. Thus, it is assumed that the DERs are operating in the grid following mode, and their controls maintain a constant active and reactive power injection into the grid. Besides, the proposed model represents dynamics of time scales from hundreds of milliseconds to tens of seconds and faster dynamics, such Electro-Magnetic Transients (EMT) are neglected.

The inverter-based generation is modeled as a voltage source converter (VSC), which is the most used converter for this purpose, using the average model. In the average model, the PWM controls are eliminated, and the output voltage is applied by an ideal voltage source using the abc-reference signal calculated by the inverter control. The power source was represented by a constant power injection in DC and the balance of power flow through the DC-link is regulated by a controller. This model is commonly used for transient stability simulations. 

The inverter has typically two control loops, the outer loop, which is responsible for calculating the currents in the d-q domain for providing the desired active and reactive output powers, and the inner control loop, which calculates the converter reference voltages in d-q domain to reach the currents calculated by the outer control. The active power is regulated to maintain the balance of power flow through the DC-link, which is ensured by regulating the voltage level at the capacitor, according to the following equation:

\begin{equation}
	i_{dref} = \left(k_{pvdc}+\dfrac{k_{ivdc}}{s}\right)\left(V_{dc}-V_{dcmeas}\right)
	\label{eq:convert}
\end{equation}
\noindent where $i_{dref}$ is the reference current in the direct axis; $K_p$ and $K_i$ are the proportional and integral controllers; Vdc is the nominal voltage of the DC-link; and $V_{dcmeas}$ is the measured voltage at the DC-link. 

Considering a typical condition for grid-connected DER, the injection of reactive power is set to zero, resulting in a quadrature current controlled to zero ($i_{qref} = 0$). 

The synchronous generator is represented by a fourth order model. It is assumed the machine is equipped with an AVR (represented by the IEEE ST1A model) and a typical governor \cite{Kundur1994}. The model presented in \cite{Kundur1994} was used for the induction motor used in the composite load model. The static part of the composite load is represented by a ZIP model, which is the combination of constant impedance, constant current, and constant power, using the following equations.  

\begin{equation}
	P=P_Z\left(\frac{V}{V_0}\right)^2+P_I\left(\frac{V}{V_0}\right)+P_P
	\label{eq:loadp}
\end{equation}

\begin{equation}
	Q=Q_Z\left(\frac{V}{V_0}\right)^2+Q_I\left(\frac{V}{V_0}\right)+Q_P
	\label{eq:loadq}
\end{equation}

\noindent where $P_Z$ and $Q_Z$ are the constant impedance active and reactive powers; $P_I$ and $Q_I$ are the constant current active and reactive powers; and $P_P$ and $Q_P$ are the constant power active and reactive powers.

\section{Approach for Parameters Identification of the Equivalent Model}
\label{sec:method}

The proposed method to estimate the parameters of the equivalent model is based on the data measured at the PCC and on two main steps, the first to evaluate the set of parameters to be estimated and the second to estimate the parameters. 

The parameters identification process is based on measurements collected at the PCC. Thus, for the training process, it is assumed that measurements of voltage magnitude ($V$), frequency ($f$), and active and reactive powers ($P$,~$Q$) for a set of disturbances are known. 

In the first step, it is investigated the sensitivity of the measured output variables in relation to each parameter of the proposed microgrid equivalent model, using the Trajectory Sensitivity Function (TSF). As pointed out by \cite{Geraldi2020}, this technique can support the evaluation of the parameters to be tuned for a specific sampling window of measurements. 

After selecting the set of parameters with significant influence in the response of the proposed equivalent model, a  simple and direct evolutionary algorithm, the DE, is applied for tuning these selected parameters from the sampled window analyzed. The measurements are used as a reference for the identification of the equivalent parameters, using as entries the system voltage and frequency and obtaining the active and reactive power from the equivalent using the estimated parameters. 

\begin{figure}[ht]
	\centering
	\includegraphics[width=0.35\textwidth]{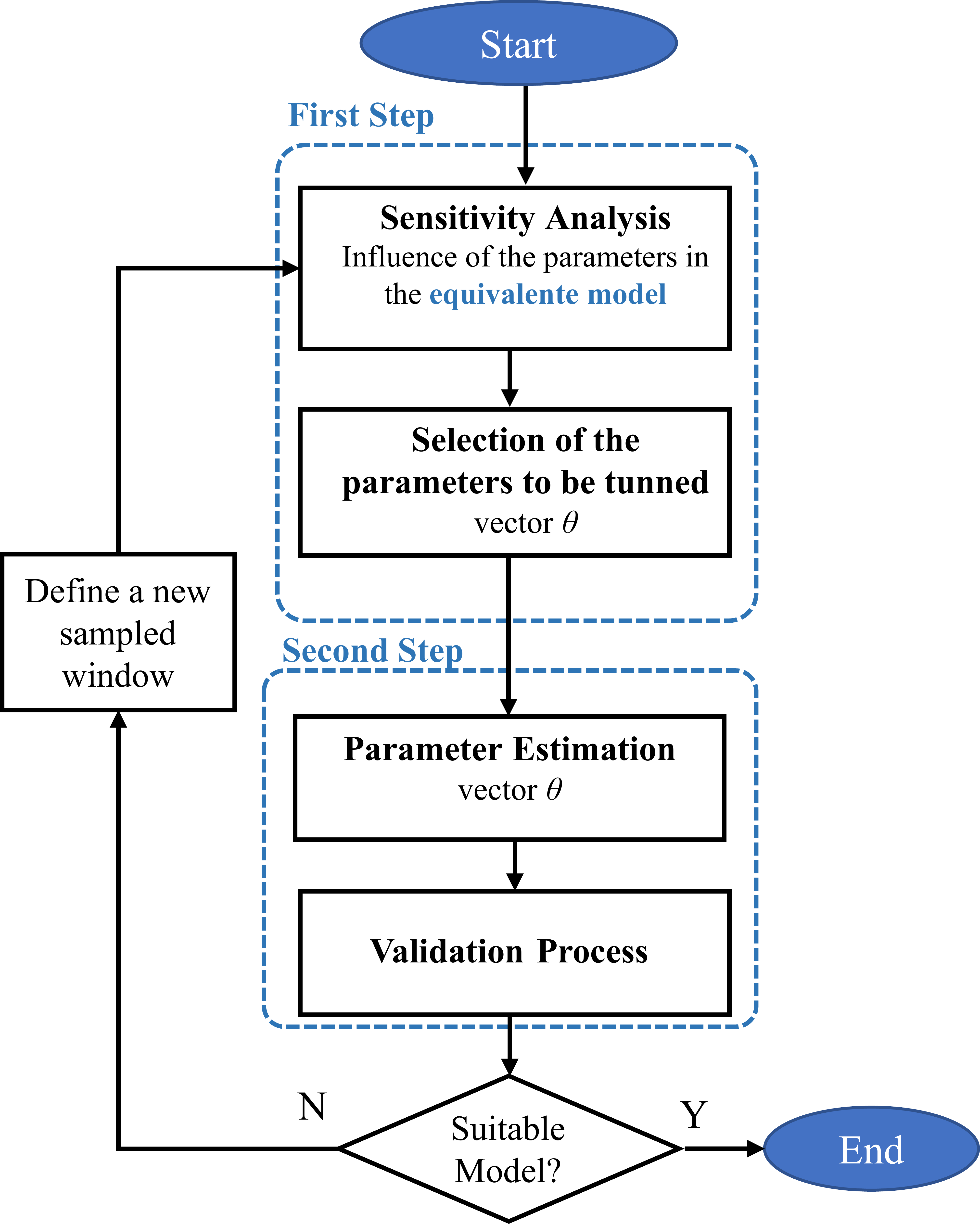}
	\caption{General overview of the approach for parameters identification of the equivalent model.}
	\label{fig:fluxogram2}
\end{figure}

An overview of the proposed approach can be seen in the Fig. \ref{fig:fluxogram2}. The details of each stage are explained in the next subsections.

\subsection{Trajectory Sensitivity Analysis}

After the structure of the equivalent model is established, it is necessary to define the set of parameters that can be estimated from the sampled data. The resulting equivalent model can be composed by a large number of parameters, depending on the set of differential algebraic equations (DAE) used to represent each element of the equivalent model. However, as mentioned by \cite{Geraldi2020}, only parameters that presents influence on the output of the equivalent model for the sampling window analyzed can be suitable tuned by the optimization method. In this sense, to eliminate the parameters that have lower influence on its dynamic the response, the trajectory sensitivity analysis is applied.

The TSF can quantify the variation of a specific trajectory resulting from small changes on the parameters \cite{Frank1978}. If a certain parameter has a large trajectory sensitivity, it indicates that this parameter have leverage in altering the model trajectory to better match the measured response, as pointed out by \cite{Hiskens2001}. On the other hand, a small value shows that a large modification in this parameter is required to significantly modify the analyzed trajectory. 

To quantify the trajectory sensitivity analysis of each parameter in relation to the outputs of the equivalent model, the index employed in \cite{Geraldi2020} is used in this work. Considering sampling window with with $N_w$ samples, this index is calculated by:

\begin{equation}
	E_{\theta_i}^{y_j} = \sum_{k=1}^{N_w} \left[\dfrac{\partial y_j(k)}{\partial \theta_i} \right]^2
	\label{eq:quadratic_func_sort}
\end{equation}

\noindent where $\theta_i$ is the $i$-th element of the parameter vector $\theta$ and $y_j$ is $j$-th output of the equivalent model. In this index, the trajectory sensitivity function of the $y_i$ regarding to $\theta_i$ for a window with $N_w$ samples is the partial derivative of $y_i$ with respect to $\theta_i$. To compute this sensitivity function, an approximate approach can also be used, as described by \cite{Geraldi2018, Frank1978}.

The index $E_{\theta_i}^{y_j}$ is calculated for all parameters in relation the both output variables of the equivalent model, $\hat{P}$ and $\hat{Q}$. The parameters that have a meaningful influence on their output response in the considering sampled window, are selected to stage 2 and grouped in $\theta$ vector. For the parameters with quite low sensitivity, the typical values can be used without affecting the final outcome.

\subsection{Estimation of the Interested Parameters}

The selected parameters are estimated using sampled data at the PCC, after the occurrence of some disturbance in the system such as a line tripping or a temporary short circuit. It is necessary that the sampled window contains the steady-state and the transient response of the system. Before performing the parameter estimation using the DE, these measurements must be pre-processed to filter out harmonics and/or noise. 

\begin{figure}[ht]
	\centering
	\includegraphics[width=0.4\textwidth]{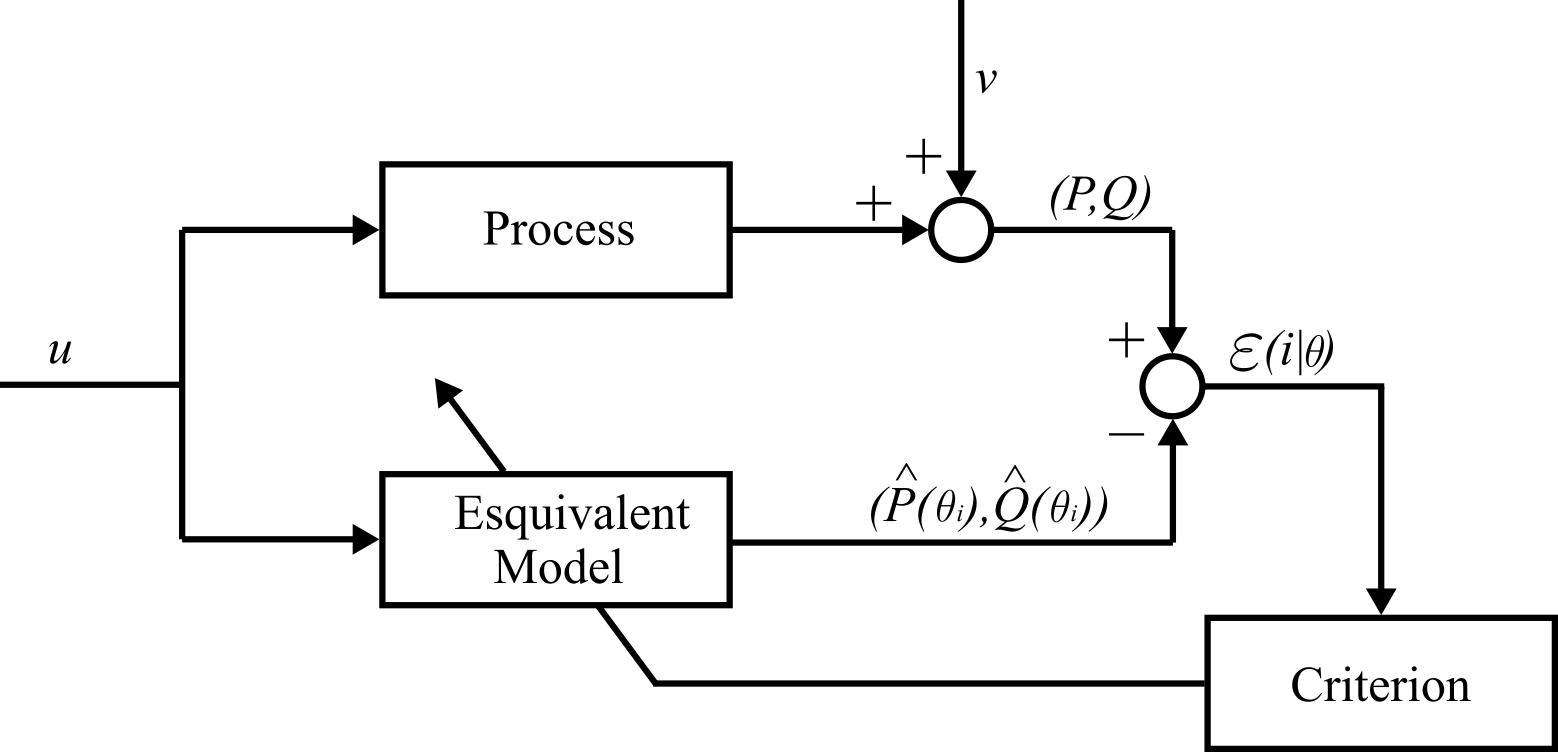}
	\caption{Parameter estimation process.}
	\label{fig:param_est}
\end{figure}

The parameter estimation is performed by the comparison between the output trajectories of the equivalent model ($\hat{P}$,~$\hat{Q}$) and the measurements taken from the PCC (${P}$,~${Q}$), with the objective of minimizing the error between the curves. For each $i$-th evaluated $\theta$ vector, containing the equivalent parameters to be tuned, the mismatch between outputs of the equivalent system ($\hat{P}(\theta_i)$,~$\hat{Q}(\theta_i)$) and the measured values are computed, as illustrated in Fig. \ref{fig:param_est}.

Similarly to other search problems, the main objective of the DE is minimizing the error between the output from the real system, obtained from measurements or from the complete system, and the outputs from the equivalent model, according to the following objective function:

\begin{align} \nonumber
	\min_{\theta}\varepsilon(\theta) =& \min_{\theta}\left[\frac{1}{N_w} \sum_{k=1}^{N_w} \left(P_k-\hat{P}_k(\theta) \right)^2\right] \\  &+\min_{\theta}\left[\frac{1}{N_w} \sum_{k=1}^{N_w} \left(Q_k-\hat{Q}_k(\theta) \right)^2\right] \\ \nonumber
	s.t \quad  &\theta^{max}\leq\theta \leq \theta^{min}
\end{align}

\noindent where, respectively, $P_k$ and $Q_k$ are the samples of the measured active power and reactive power, and $\hat{P_k}(\theta)$ and $\hat{Q_k}(\theta)$ are the samples of the simulated active power and reactive power using the equivalent grid and the parameters of the vector $\theta$. The bounds $\theta^{max}$ and  $\theta^{min}$ allows to define the search region, taking into account typical values of each parameter of the equivalent model. In relation to the meta-heuristic
optimization method DE and its particularities, additional information can be seen in \cite{Kenneth2005}.

After the parameter estimation, to evaluate the performance of the proposed equivalent model with the tuned parameters, the match between the output signal provided by the equivalent model and the output of the real system is calculated by the mean squared error (MSE): 

\begin{equation}
	MSE = \frac{1}{N_{w}} \sum_{k=1}^{N_{w}}(y_k - \hat{y}_{k})^2
\end{equation}
\noindent where $\hat{y}_k$ is the output given by equivalent model $k$, the response of the real system is denoted by $y_{k}$. 

If the error between both models is not significant, the validation process is applied. In this process, it is verified whether the equivalent model with tuned parameters is capable to reproduce (or predict) adequately the dynamic behavior of the real system for different disturbance events.
	
However, if the equivalent model is not validated, the first and second step must be repeated. With this purpose, a new sampled window that contains the response of the system to a different disturbance could be used for the estimation process. As pointed out by \cite{Geraldi2020}, it is also possible to use the same disturbance applied before, but it is required a new time window of the measured signals, resulting in a different proportion between the steady-state and transient response.
		
\section{Simulation Results and Discussion}
\label{sec:results}
	
\begin{figure}[ht]
	\centering
	\includegraphics[width=0.48\textwidth]{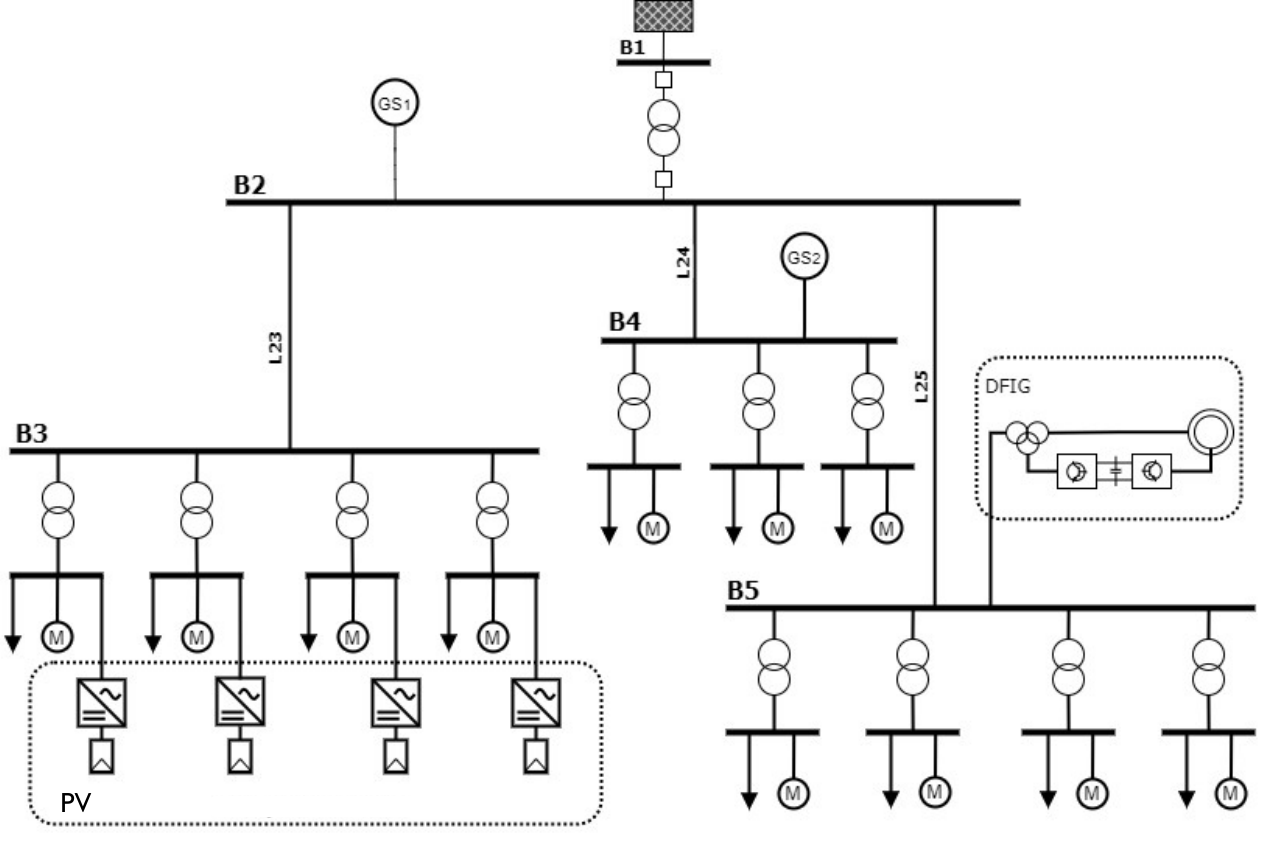}
	\caption{Test system.}
	\label{fig:test_system}
\end{figure}

To verify the effectiveness of the proposed method to build equivalent models for a microgrid, the test system presented in Fig. \ref{fig:test_system} is used. This system is a modified version of the system presented in \cite{Milanovic2006} and represents part of a typical medium-voltage distribution system. The microgrid sources are composed by two synchronous generators, two wind generators, and PV systems. The synchronous generators are driven by gas turbine units. The first one presents a capacity of 6.9 MW and is connected to bus 02, while the second one has a capacity of 1.9 MW and is connected to bus 4. The AVRs of both generators are modeled by a simplified IEEE ST1A model and they operate at unity power factor. A total capacity of 500 kW of PV generation is considered in the system. These PV unitus are equally distributed across 4 feeders (close to the loads) to represent the PV generation connected at low-voltage level distribution systems. Two wind-generators are also considered: a fixed-speed wind turbine of 1.5 MW, connected to bus 3, and a doubly fed induction generator-based wind turbine of 1.5 MW, connected to bus 5. 
		
The system is modeled in MATLAB/Simulink, considering the Electro-Magnetic Transients (EMT) models of all microgrid elements. The boundary between the transmission and the microgrid is defined by bus 01, at the PCC. The microgrid elements, downstream of this border, are replaced by the proposed equivalent model.
		
To obtain the input and output signals for estimation the dynamic equivalent model, a three-phase short circuit is applied between the PCC (bus 01) and the substation in the test system, during a interval of 500~ms, at time $t$=10~s with a fault resistance of 2~$\Omega$. The measurements of active and reactive power responses at the PCC, as well as the voltage and frequency at the same bus are recorded with a sampling rate of 10 ms, in a time window of 9-14s. 
		
The equivalent system has also been implemented in MATLAB/Simulink, where a three-phase round rotor is used to represent the SG, by a seventh-order state-space model \cite{Kundur1994}. The SG is equipped with a AVR modeled by a simplified IEEE ST1A model (as the test system). The induction motor is a squirrel-cage machine, represented by sixth-order state-space model \cite{Krause2002}. The static part of the composite load and the VSC are represented by the equations described in section \ref{sec:equivalent_model}.  The idea is to use  typical values of parameters as initial values for the estimation, thus we used the values of parameters generator and loads used in \cite{Milanovic2013} for the ADN presetend in this reference.
		
\subsection{Selection of the Parameters to be Estimated }
		
As presented in section \ref{sec:method}, the sensitivity analysis  was applied to support the selection of the parameters to be tuned. With this purpose, the same three-phase short circuit applied in the test system is also simulated in the equivalent model. The measures of  ($\hat{P}$,~$\hat{Q}$) at the PCC, in the time window of 9-14s are taken as outputs.
		
		
Table \ref{tab:sensibilidade} shows the quadratic function values of TSF for each parameter of the SG model (seventh-order model) with respect to each output, in the period pre-disturbance (time window between 9-9.99s) and a sampled window containing the response of this disturbance (10-14s), pos-disturbance. In the pre-disturbance period, the parameters $x_d,~x_q,~T{'}_{do}$ achieve the highest index values in both measured signals (in bold in the Table \ref{tab:sensibilidade}). 
		
On the other hand, the rank of parameters taking into account $E^{\hat{P}}_{\theta_i}$ is quite different from the one obtained considering $E^{\hat{Q}}_{\theta_i}$ in the transient portion of the response. In this way, the combination of the quadratic functions calculated from both signals was used to obtain the raking of the parameters. From this analysis, the parameters $H,~x{'}_d,~x{'}_q$ were selected to be estimated. To complete the set of parameters that represents a fourth-order model of SG, the parameter $T^{'}_{q}$ is also included in this selection.
		
\vspace{-0.3cm}
		
\begin{center}
	\begin{table}[!ht]
		\centering
		\caption{Quadratic Function of each parameter $\theta_i$.}
		\resizebox{.47\textwidth}{!}{
			\begin{tabular}{c  c c  c c}
				\hline
				\textbf{Parameter} &\multicolumn{2}{c}{\textbf{Pre-Disturbance}}  &\multicolumn{2}{c}{\textbf{Pos-Disturbance}} \\
						
				\textbf{$\theta_i$} &\textbf{$E^{\hat{P}}_{\theta_i}$} &\textbf{$E^{\hat{Q}}_{\theta_i}$} &\textbf{$E^{\hat{P}}_{\theta_i}$} &\textbf{$E^{\hat{Q}}_{\theta_i}$}  \\
				\hline
				$x_d$	    &\textbf{3.23$\times$10$^{-4}$}	&\textbf{5.32$\times$10$^{-1}$}	&\textbf{2.57$\times$10$^{-2}$}	&\textbf{2.66} \\
				$x'_{d}$	    &9.48$\times$10$^{-11}$	            &3.05$\times$10$^{-7}$	             &5.34$\times$10$^{-2}$   &\textbf{2.14}  \\
				$x''_{d}$	&6.74$\times$10$^{-14}$          	&1.84$\times$10$^{-10}$	           &4.92$\times$ 10$^{-3}$	  &9.70$\times$10$^{-2}$ \\
				$x_q$	       &\textbf{1.87$\times$10$^{-7}$}	&\textbf{1.33$\times$10$^{-4}$}	&4.75$\times$10$^{-3}$	&1.54$\times$10$^{-2}$ \\
				$x'_{q}$	&1.89$\times$10$^{-9}$	&6.34$\times$10$^{-8}$	&5.26$\times$10$^{-2}$	&7.52$\times$10$^{-2}$ \\
				$x''_{q}$	&1.91$\times$10$^{-10}$	&6.15$\times$10$^{-9}$	&\textbf{8.76$\times$10$^{-2}$}	&1.17$\times$10$^{-1}$ \\
				$x_l$	&9.41$\times$10$^{-22}$	&1.53$\times$10$^{-20}$	&3.10$\times$10$^{-21}$	&4.54$\times$10$^{-20}$ \\
				$T'_{do}$	&\textbf{2.54$\times$10$^{-9}$}	&\textbf{1.81$\times$10$^{-6}$}	&3.22$\times$10$^{-4}$	&2.56$\times$10$^{-2}$ \\
				$T''_{do}$	&1.80$\times$10$^{-14}$	&3.57$\times$10$^{-11}$	&5.13$\times$10$^{-4}$	&1.12$\times$10$^{-2}$ \\
				$T_{q}$	&8.97$\times$10$^{-11}$	&3.15$\times$10$^{-09}$	&2,72$\times$10$^{-4}$	&3.37$\times$10$^{-4}$ \\
				$T'_{q}$	&1.00$\times$10$^{-10}$	&3.20$\times$10$^{-9}$	&3.51$\times$10$^{-2}$	&4.23$\times$10$^{-2}$ \\
				$H$	&8.79$\times$10$^{-10}$	&4.99$\times$10$^{-10}$	&\textbf{1.11}	&\textbf{3.28$\times$10$^{-1}$} \\
				\hline
		\end{tabular}}
		\label{tab:sensibilidade}
	\end{table}
\end{center} 

\vspace{-0.3cm}
	
For the other elements of the microgrid, the same procedure is applied. It is selected the smallest set of parameters to represent each component, taking into account the best trade-off between  precision, computational burden, and trajectory sensitivity. Table \ref{tab:parameters} summarizes the parameters to be identified by the optimization method (total of 20 parameters), where $K_a$ is the gain of AVR; $S_m$,~$x_m$,~$H_m$ are is the nominal apparent power, the mutual reactance and the inertia constant of the induction motor, respectively; and, $S_{VSC}$ is the nominal apparent power of the voltage source converter.

		
\vspace{-0.3cm}
		
\begin{center}
	\begin{table}[!ht]
		\centering
		\caption{Estimated parameters of the equivalent model.}
			\begin{tabular}{c c }
				\hline
				\textbf{Element Model} &\textbf{Parameter}  \\
				\hline
				Synchronous generator  &$X_d$,~$X_q$,~$X'_d$,~$X'_q$,~$T'_{do}$, ~$T'_{q}$,~$H$\\
				AVR     & $K_a$\\ 
				Voltage source converter  &$S_{VSC}$,~$K_{ivd}$,~$K_{pvdc}$\\
				Static Load  &$P_{P}$,~$Q_{P}$,~$P_{I}$,~$Q_{I}$,~$P_{z}$,~$Q_{z}$\\
				Induction Motor  &$S_m$,~$H_m$,~$X_{m}$ \\
				\hline
			\end{tabular}
			\label{tab:parameters}
	\end{table}
\end{center} 
			
\subsection{Parameter Estimation}
		
From the input and output signals of the complete model measured at PCC, the parameters are estimated in two stages. The pre-disturbance period (time window between 9-9.99s) is used to estimate a first group of parameters, whereas, for the second parameters group, the full test system response to this disturbance is required (time window between 10-14s). In both groups, the search region is limited to the ranges of typical values of the parameters of synchronous machines and control parameters of inverter-based generation, which are presented in Table  \ref{tab:TypicalRange}. Furthermore, it is assumed that the total capacity of the synchronous generators connected to the microgrid is known, since this type of data is usually easily available for the utility companies. The remain parameters of each equivalent system are fixed in their reference values, which were imported from the full test system.
			
In the first stage, the parameters related to load power components are selected, since they contribute to the power balance and, consequently, reached a significantly sensitivity in the steady-state. Furthermore, it is also included the SG parameters $x_d,~x_q$, and $T'_{do}$, based on their sensitivity analysis when the both output signals are considered, as can be seen in Table \ref{tab:sensibilidade}. 
			
\vspace{-0.3cm}
			
\begin{center}
	\begin{table}[!ht]
		\centering
		\caption{Ranges of typical values of the parameters.}
		\resizebox{.47\textwidth}{!}{
			\begin{tabular}{c c c c c c}
				\hline
				\textbf{$\theta_i$}  &\textbf{Value} &\textbf{$\theta_i$} &\textbf{Value} &\textbf{$\theta_i$}  &\textbf{Value}  \\
				\hline
				$x_d$ (pu)   &$[$0.5 - 3.0$]$   &$T'_{do}$ (s)	&$[$0.5 - 8.0$]$   &$K_{pvdc}$ &$[$0.1 - 2.0$]$ \\
				$x'_d$ (pu)   &$[$0.05 - 0.5$]$   &$T'_{q}$ (s)	&$[$0.5 - 2.0$]$ &$K_{ivdc}$ &$[$20 - 500$]$ \\
				$x_q$ (pu)   &$[$0.5 - 3.0$]$   &$H$ (s)	&$[$0.01 - 5.0$]$  &$K_a$ &$[$50-400$]$ \\
				$x'_q$ (pu)   &$[0.3 - 1.0]$    &$T_a$ (s) &$[$0.0001 - 0.01$]$  &   &\\
				\hline
			\end{tabular}}
		\label{tab:TypicalRange}
	\end{table}
\end{center}
			
\vspace{-0.3cm}
			
Thus, the resulting parameter vector is given by:  
			
\footnotesize
\begin{equation} \nonumber
	\theta_{pre}=\left[~P_{P},~Q_{P},~P_{I},~Q_{I},~P_{z},~Q_{z},~S_{m},~S_{VSC},~x_d,~x_q,~T'_{do},\right].   
\end{equation} 
\normalsize
			
\vspace{0.1cm}
			
The DE population is initialized by a random generation of 15 individuals, where each individual is a population vector. A mutation scaling factor ($F_s$) of 0.8 and a crossover constant ($C_r$) of 0.3 are used in the algorithm. Table \ref{tab:limites} shows the search region for each parameter related to the load power. The initial values of the SG parameters were considered as known values, assuming a search region from 80\% to 120\%. 
			
\vspace{-0.3cm}
			
\begin{center}
	\begin{table}[!ht]
		\centering
		\caption{Search region of the parameters related to the load power in the first stage.}
			\begin{tabular}{c c c c c }
				\hline
				\textbf{Limits}  &\textbf{$P_Z,~P_I,~ P_P$} &\textbf{$Q_Z,~Q_I,~Q_P$} &$S_m$    &$S_{VSC}$\\
				\hline
				$\theta^{min}_i$    &1 MW	 &0.2 MVar &0.5 MVA  &2.4 MVA \\
				$\theta^{max}_i$    &3 MW  &2 MVar &1.5 MVA  &3.6 MVA\\
				\hline
			\end{tabular}
			\label{tab:limites}
	\end{table}
\end{center} 
				
\vspace{-0.3cm}
				
In the second stage, taking into account the values of the parameters that had already been tuned, the remaining parameters are estimated. For this step, the population of DE is composed of 30 individuals, $F_s$ and $C_r$ are settled as 0.8 and 0.7, respectively. The initial values of each parameter are uniformly chosen at random between an interval, from 80\% to 120\% of their reference values.
				
\vspace{-0.3cm}
				
\begin{center}
	\begin{table}[ht]
		\centering
		\caption{Parameters estimated for the equivalent model.}
		\begin{tabular}{c c c c c c}
			\hline
			\textbf{$\theta_i$}  &\textbf{Value} &\textbf{$\theta_i$} &\textbf{Value} &\textbf{$\theta_i$}  &\textbf{Value}  \\
			\hline
			$x_d$   &2.633 pu   &$K_a$	&177.995  &$Q_I$ &1.517 MVar\\
			$x'_d$   &0.282 pu   &$S_m$	&1.152 MVA &$P_Z$ &1.154 MW\\
			$x_q$   &1.600 pu   &$H_m$	&0.550 s  &$Q_Z$ &1.327 MVar\\
	 		$x'_q$   &0.964 pu   &$X_m$	&2.001 pu  &$S_{VSC}$ &3.027 MVA\\
			$T'_{do}$   &6.76 pu   &$P_P$	&2.536 MW  &$K_{pvdc}$ &1.636\\
			$T'_{q}$   &0.914 pu   &$Q_P$	& 0.978 MVar  &$K_{ivdc}$ &457.07\\
			$H$   &3.108 s   &$P_I$	&1.512 MW  &   &\\
			\hline
		\end{tabular}
		\label{tab:ParametersEstimated}
	\end{table}
\end{center} 
				
\vspace{-0.3cm}
				
\begin{figure}[ht]
	\centering
	\includegraphics[width=0.48\textwidth]{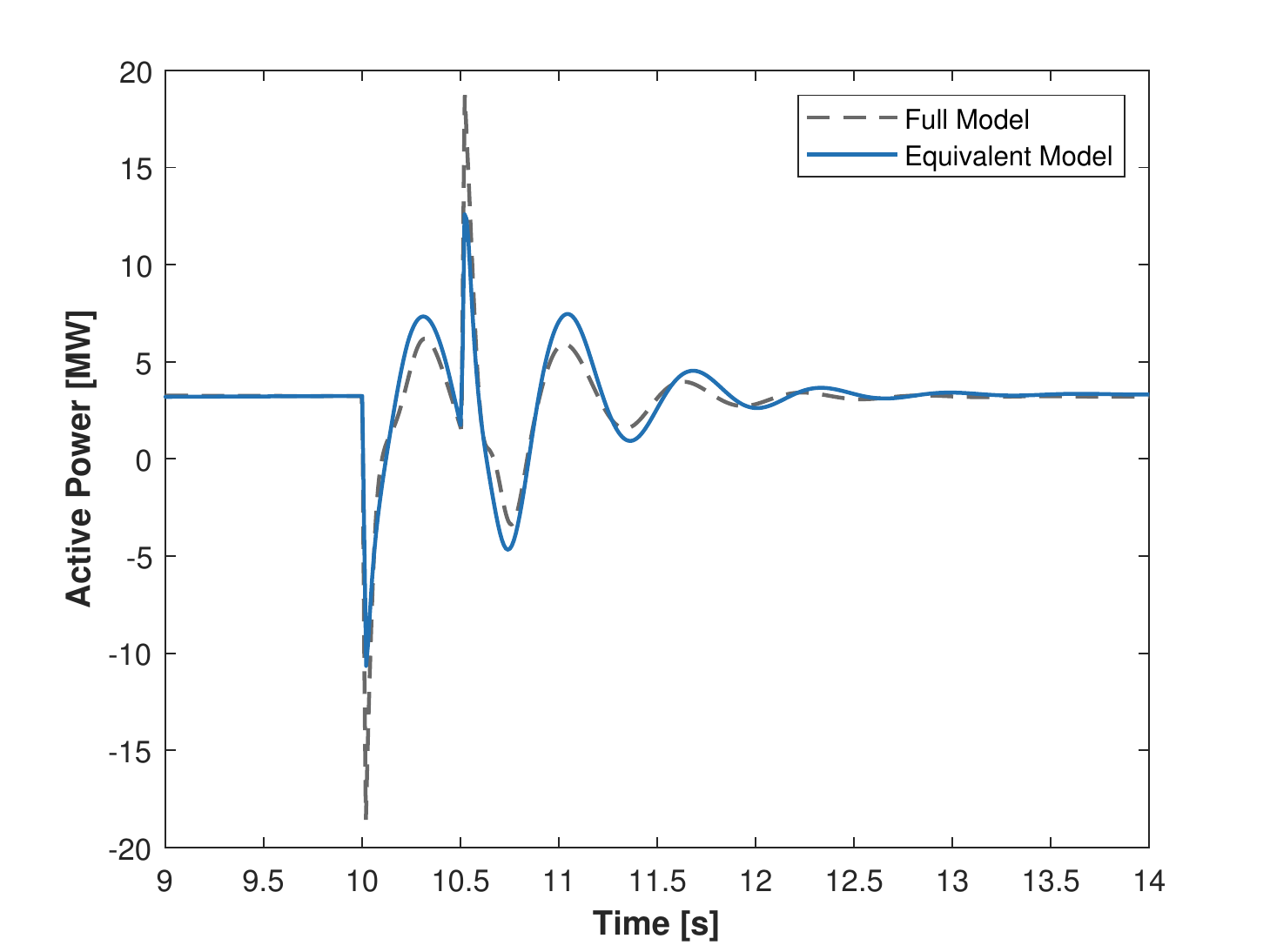}
	\caption{Comparison between active power response for the equivalent model and full model of the microgrid.}
	\label{fig:active_power}
\end{figure}

The active power response from the full model in comparison with the proposed equivalent model is exhibited in Fig. \ref{fig:active_power}. As can be seen, the behaviors during and after the disturbance are very similar. 
				
\begin{figure}[ht]
	\centering
	\includegraphics[width=0.48\textwidth]{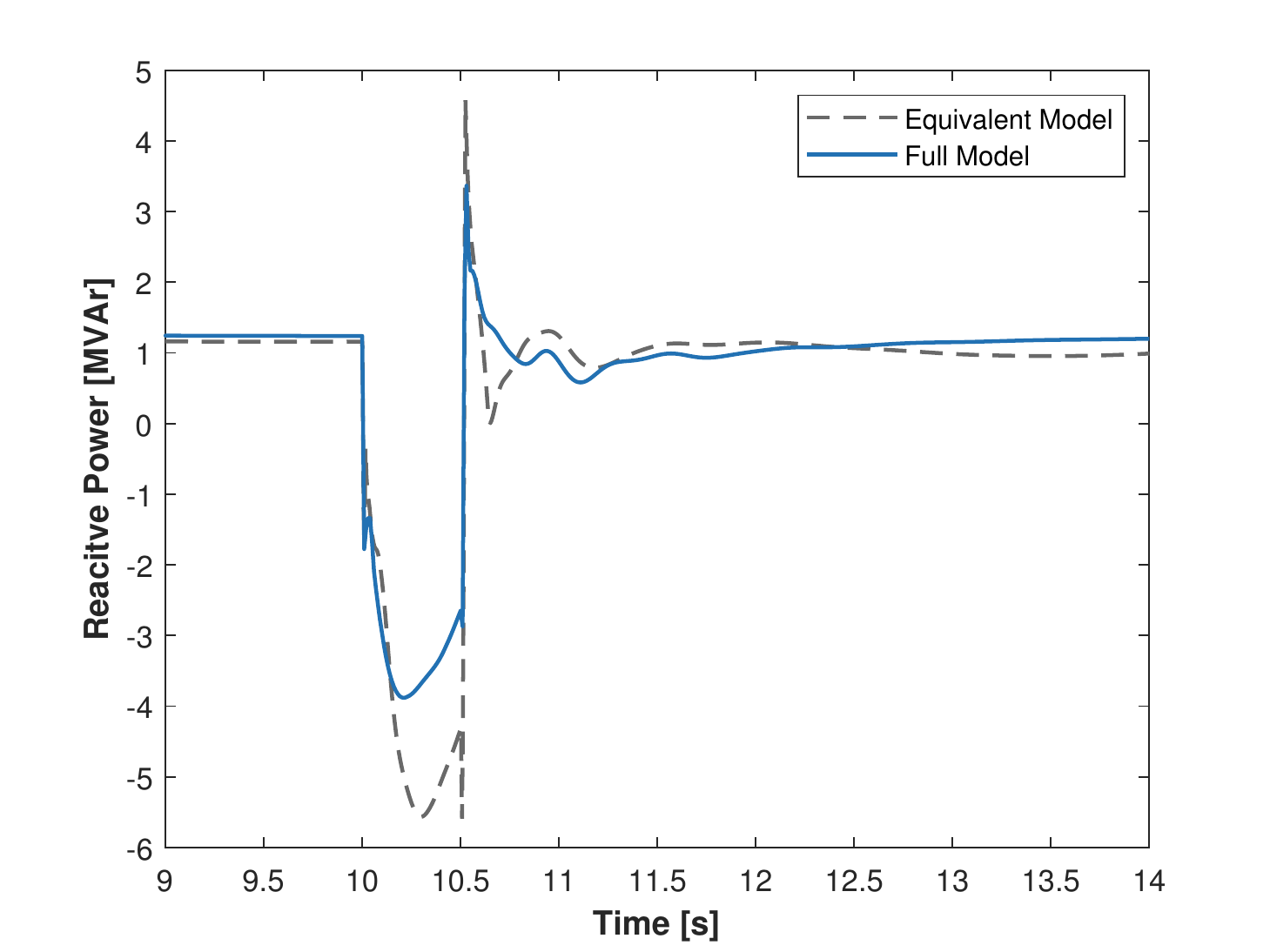}
	\caption{Comparison between reactive power response for the equivalent model and full model of the microgrid.}
	\label{fig:reactive_power}
\end{figure}

Fig. \ref{fig:reactive_power} presents the reactive power also from the full model and the equivalent model, which shows that the equivalent model presents a good match, from the qualitative viewpoint (which is the main purpose of an equivalent model), with the full model for the power exchanges.
				
\begin{figure}[ht]
	\centering
	\includegraphics[width=0.48\textwidth]{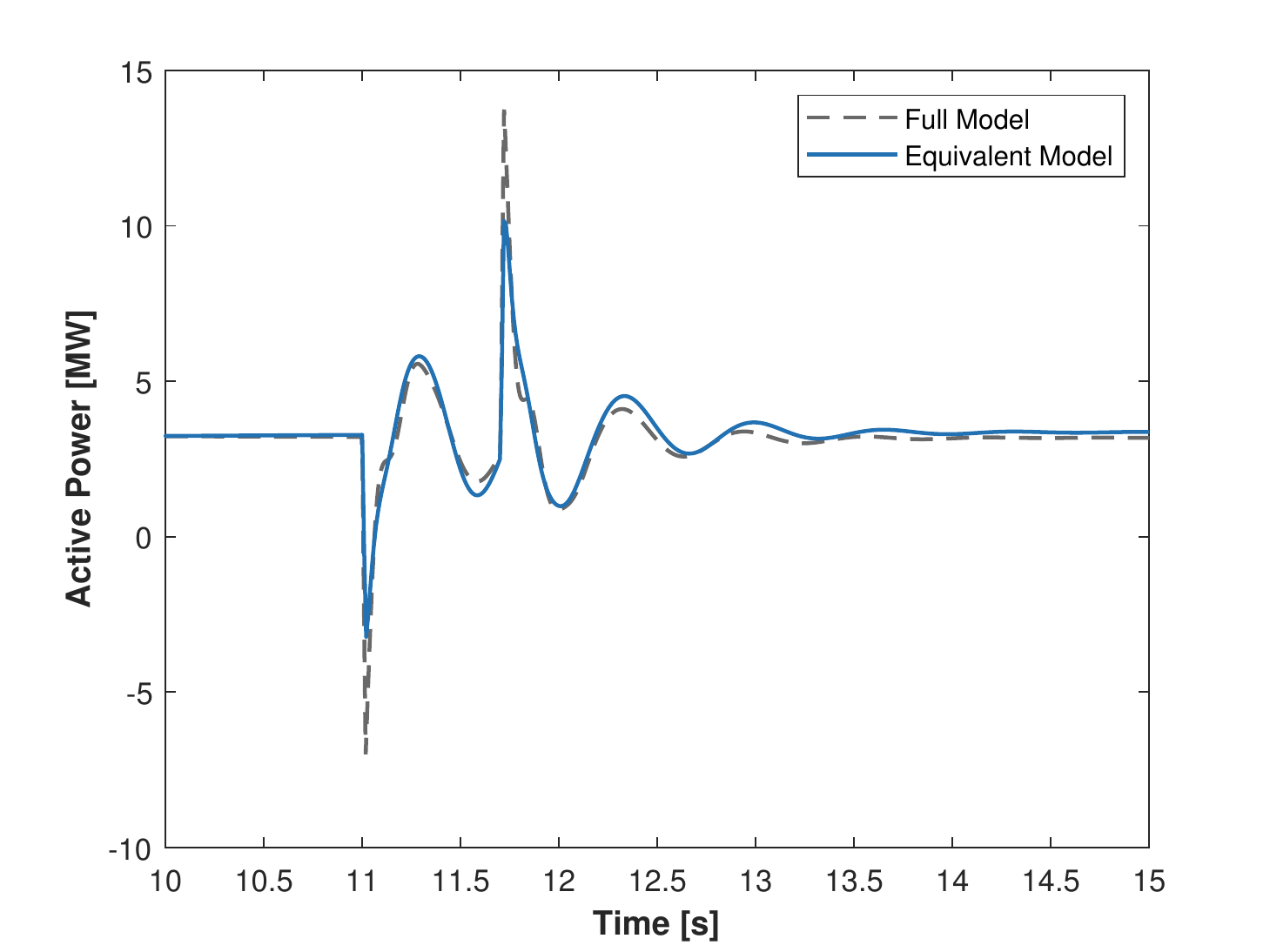}
	\caption{Comparison between active power response for the equivalent model and full model of the microgrid in the validation.}
	\label{fig:active_power_val}
\end{figure}

\begin{figure}[ht]
	\centering
	\includegraphics[width=0.48\textwidth]{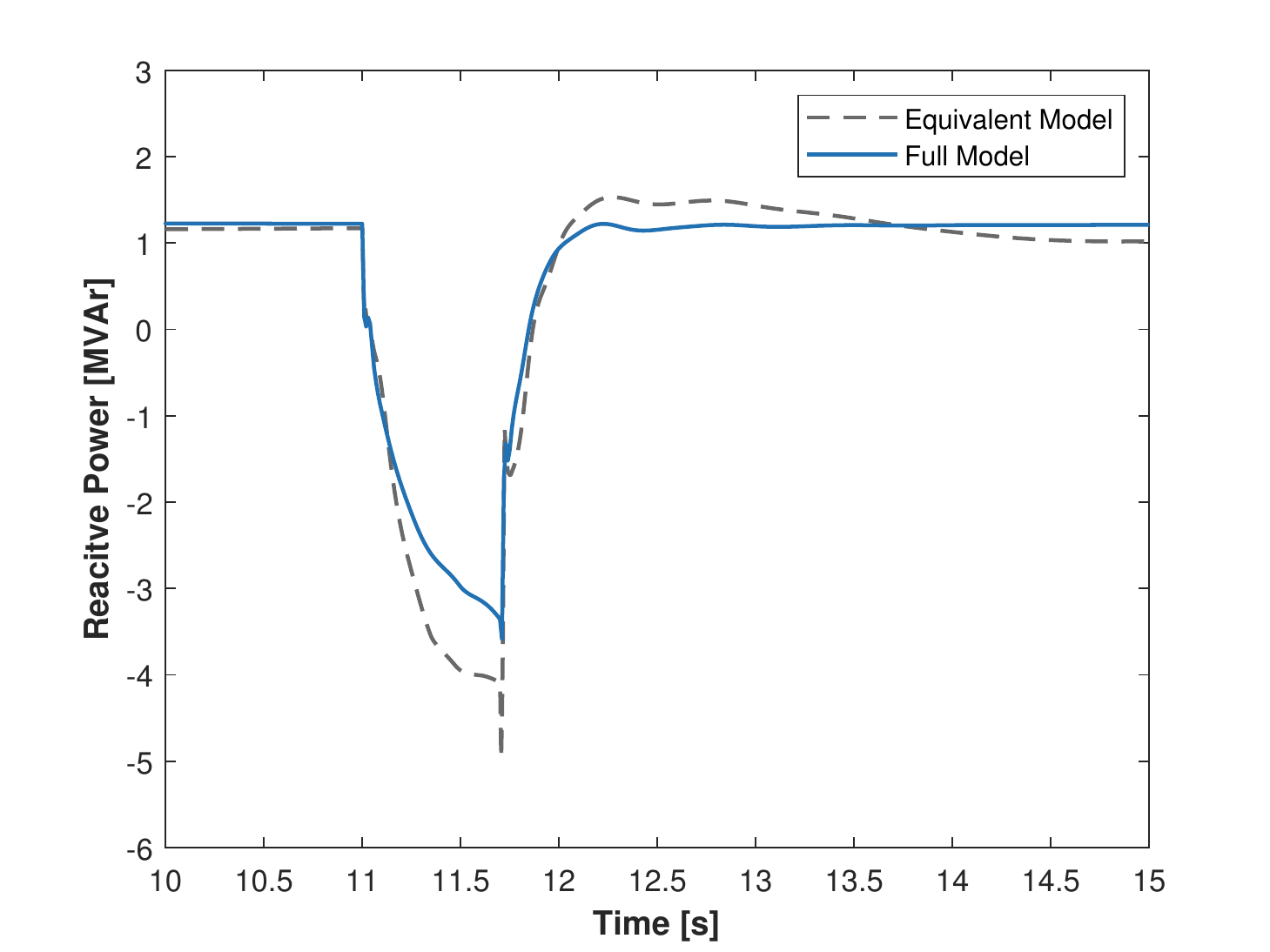}
	\caption{Reactive power response for the equivalent model and full model of the microgrid in the validation.}
	\label{fig:reactive_power_val}
\end{figure}

To validate the equivalent system with the estimated parameters, the response of the full system to a new disturbance is recorded. The disturbance applied consists of a three-phase fault between the PCC and the substation, at the instant 11.0 s, with a duration of 700 ms. The fault resistance is equal to 5.0 $\Omega$.  The comparison between the responses of the estimated equivalent model and the full model are shown in Fig. \ref{fig:active_power_val} and Fig. \ref{fig:reactive_power_val}, respectively, and the observed qualitative matches provide the desired validation.
				
The MSE between the output signals of the full model and the equivalent model for both disturbances is shown in Table \ref{tab:MSE}. Note that the performance of the equivalent model is adequate in both situations.
				
\vspace{-0.3cm}
				
\begin{center}
	\begin{table}[ht]
		\centering
		\caption{MSE calculated for each fault evaluated.}
			\begin{tabular}{c c c c c}
				\hline
				\textbf{Fault}  &\textbf{Window} &\textbf{$MSE_P$} &\textbf{$MSE_Q$} &\textbf{$MSE_{P+Q}$}  \\
				\hline
				Fault 1   & 9-14s   &0.729	&0.246    &1.038  \\
				Fault 2  & 10-15s  &0.157   &0.113   &0.270 \\
				\hline
			\end{tabular}
			\label{tab:MSE}
		\end{table}
\end{center} 
					
\vspace{-0.5cm}
					
				
\section{Conclusions}
\label{sec:conclusions}
				
This paper presented an equivalent, based on a gray-box approach, to represent grid-connected microgrids or active distribution networks in dynamic studies. The equivalent is able to represent a typical mix of generation that can be connected to distribution systems, as well as the load behavior. The equivalent based on gray-box approach is practical to be integrated in dynamic simulation tools, as it is based on typical dynamic components available in these programs. 
				
The main contribution of the method is the use of trajectory sensitivity analysis to select the most important parameters to be estimated, thus reducing the dimension of the search space. This proposal allowed the use of a simple search algorithm to estimate the parameters of the equivalent model, increasing its applicability to the industry practice. The results confirmed that the equivalent is able to represent the dynamic response of an active distribution network or microgrid and can be used to generate computationally feasible simulations for dynamic analyses of power systems. 

\section*{Acknowledgment}

The authors would like to thank the financial support of Fapesp (under Grant \#2018/20104-9) and CNPq (under Grant \#43865/2018-6).



\bibliographystyle{IEEEtran}
\bibliography{IEEEabrv,references}

\begin{thebibliography}{10}
\providecommand{\url}[1]{#1}
\csname url@samestyle\endcsname
\providecommand{\newblock}{\relax}
\providecommand{\bibinfo}[2]{#2}
\providecommand{\BIBentrySTDinterwordspacing}{\spaceskip=0pt\relax}
\providecommand{\BIBentryALTinterwordstretchfactor}{4}
\providecommand{\BIBentryALTinterwordspacing}{\spaceskip=\fontdimen2\font plus
\BIBentryALTinterwordstretchfactor\fontdimen3\font minus
  \fontdimen4\font\relax}
\providecommand{\BIBforeignlanguage}[2]{{%
\expandafter\ifx\csname l@#1\endcsname\relax
\typeout{** WARNING: IEEEtran.bst: No hyphenation pattern has been}%
\typeout{** loaded for the language `#1'. Using the pattern for}%
\typeout{** the default language instead.}%
\else
\language=\csname l@#1\endcsname
\fi
#2}}
\providecommand{\BIBdecl}{\relax}
\BIBdecl

\bibitem{Olivares2014}
D.~E. Olivares, A.~Mehrizi-Sani, A.~H. Etemadi, C.~A. Canizares, R.~Iravani,
  M.~Kazerani, A.~H. Hajimiragha, O.~Gomis-Bellmunt, M.~Saeedifard,
  R.~Palma-Behnke, G.~A. Jimenez-Estevez, and N.~D. Hatziargyriou, ``{Trends in
  Microgrid Control},'' \emph{IEEE Trans. Smart Grid}, vol.~5, no.~4, pp.
  1905--1919, jul 2014.

\bibitem{Milanovic2013}
J.~V. Milanovic and S.~{Mat Zali}, ``{Validation of Equivalent Dynamic Model of
  Active Distribution Network Cell},'' \emph{IEEE Trans. Power Syst.}, vol.~28,
  no.~3, pp. 2101--2110, aug 2013.

\bibitem{Zaker2019}
B.~Zaker, G.~B. Gharehpetian, and M.~Karrari, ``{A Novel Measurement-Based
  Dynamic Equivalent Model of Grid-Connected Microgrids},'' \emph{IEEE Trans.
  Ind. Informatics}, vol.~15, no.~4, pp. 2032--2043, apr 2019.

\bibitem{resende}
F.~O. Resende and J.~A.~P. Lopes, ``Development of dynamic equivalents for
  microgrids using system identification theory,'' in \emph{2007 IEEE Lausanne
  Power Tech}, 2007, pp. 1033--1038.

\bibitem{erlich2004a}
A.~M. Azmy, I.~Erlich, and P.~Sowa, ``Artificial neural network-based dynamic
  equivalents for distribution systems containing active sources,'' in
  \emph{Proc. Inst. Elect. Eng.}, vol. 151, 2004, p. 681–688.

\bibitem{erlich2004b}
------, ``Artificial neural network-based dynamic equivalents for distribution
  systems containing active sources,'' \emph{IEE Proc. Gener. Transm.
  Distrib.}, vol. 151, no.~6, p. 681–688, 2004.

\bibitem{pana2015}
P.~N. Papadopoulos, T.~A. Papadopoulos, P.~Crolla, A.~J. Roscoe, G.~K.
  Papagiannis, and G.~M. Burt, ``Measurement-based analysis of the dynamic
  performance of microgrids using system identification,'' \emph{IET Gener.
  Transm. Distrib.}, vol.~9, no.~1, pp. 90--103, 2015.

\bibitem{chasp2018}
G.~Chaspierre, P.~Panciatici, and T.~Van~Cutsem, ``Aggregated dynamic
  equivalent of a distribution system hosting inverter-based generators,'' in
  \emph{2018 Power Systems Computation Conference (PSCC)}, 2018, pp. 1--7.

\bibitem{chasp2020}
G.~Chaspierre, G.~Denis, P.~Panciatici, and T.~Van~Cutsem, ``An active
  distribution network equivalent derived from large-disturbance simulations
  with uncertainty,'' \emph{IEEE Transactions on Smart Grid}, vol.~11, no.~6,
  pp. 4749--4759, 2020.

\bibitem{conte2019}
F.~Conte, F.~D'Agostino, and F.~Silvestro, ``{Operational constrained nonlinear
  modeling and identification of active distribution networks},'' \emph{Electr.
  Power Syst. Res.}, vol. 168, pp. 92--104, mar 2019.

\bibitem{nuno2020}
N.~Fulg{\^{e}}ncio, C.~Moreira, L.~Carvalho, and J.~{Pe{\c{c}}as Lopes},
  ``{Aggregated dynamic model of active distribution networks for large voltage
  disturbances},'' \emph{Electr. Power Syst. Res.}, vol. 178, p. 106006, jan
  2020.

\bibitem{Kundur1994}
P.~Kundur, \emph{Power System Stability and Control}.\hskip 1em plus 0.5em
  minus 0.4em\relax New York, NY: McGraw-Hill, 1994.

\bibitem{Geraldi2020}
E.~L. Geraldi, T.~C. Fernandes, A.~B. Piardi, A.~P. Grilo, and R.~A. Ramos,
  ``Parameter estimation of a synchronous generator model under unbalanced
  operating conditions,'' \emph{Electric Power Systems Research}, vol. 187, p.
  106487, 2020.

\bibitem{Frank1978}
P.~M. Frank, \emph{Introduction to System Sensitivity Theory}.\hskip 1em plus
  0.5em minus 0.4em\relax New York: Academic Press, 1978.

\bibitem{Hiskens2001}
I.~Hiskens, ``Nonlinear dynamic model evaluation from disturbance
  measurements,'' \emph{IEEE Transactions on Power Systems}, vol.~16, no.~4,
  pp. 702--710, 2001.

\bibitem{Geraldi2018}
E.~L. Geraldi, T.~C. Fernandes, and R.~A. Ramos, ``A {UKF}-based approach to
  estimate parameters of a three-phase synchronous generator model,''
  \emph{Energy Systems}, vol.~9, pp. 573--603, 2018.

\bibitem{Kenneth2005}
K.~V. P. M. S. J.~A. Lampinen, \emph{Differential Evolution: A Practical
  Approach to Global Optimization}.\hskip 1em plus 0.5em minus 0.4em\relax
  Berlin, Heidelberg: Springer, 2005.

\bibitem{Milanovic2006}
J.~V. Milanovic and M.~Kayikci, ``Transient responses of distribution network
  cell with renewable generation,'' in \emph{2006 IEEE PES Power Systems
  Conference and Exposition}, 2006, pp. 1919--1925.

\bibitem{Krause2002}
P.~Krause, O.~Wasynczuk, and S.~Sudhoff, \emph{Analysis of Electric
  Machinery}.\hskip 1em plus 0.5em minus 0.4em\relax IEEE Press, 2002.

\end{thebibliography}
%
%
%

\end{document}